\documentclass[
 reprint,
 amsmath,amssymb,
 aps,prl
]{revtex4-2}

\usepackage{graphicx}
\usepackage{dcolumn}
\usepackage{bm}
\usepackage{xcolor}
\usepackage{soul}
\usepackage{mathtools} 
\usepackage[hidelinks]{hyperref}
\usepackage{booktabs}

\makeatletter
\def\maketitle{
\@author@finish
\title@column\titleblock@produce
\suppressfloats[t]}
\makeatother

\begin{document}

\preprint{APS/123-QED}

\title{Hybrid satellite-fiber quantum network}

\author{Yanxuan Shao$^{1,2}$}
\author{Saikat Guha$^{3,4,5}$}
\author{Adilson E. Motter$^{1,2,6,7,8}$}
\affiliation{%
$^1$Department of Physics and Astronomy, Northwestern University, Evanston, IL 60208 \\
$^2$Center for Network Dynamics, Northwestern University, Evanston, IL 60208\\
$^3$James C. Wyant College of Optical Sciences, University of Arizona, Tucson, AZ 85721\\
$^4$Department of Electrical and Computer Engineering, University of Arizona, Tucson, AZ 85721\\
$^5$Electrical and Computer Engineering, University of Maryland, College Park, MD 20742\\
$^6$Northwestern Institute on Complex Systems, Northwestern University, Evanston, IL 60208\\
$^7$Department of Engineering Sciences and Applied Mathematics, Northwestern University, Evanston, IL 60208 \\
$^8$Institute for Quantum Information Research and Engineering, Northwestern University, Evanston, IL 60208
}

\date{\today}

\begin{abstract}
\noindent
Quantum networks hold promise for key distribution, private and distributed computing, and quantum sensing, among other applications. The scale of such networks for ground users is currently limited by one's ability to distribute entanglement between distant locations. This can in principle be carried out by transmitting entangled photons through optical fibers or satellites. The former is limited by fiber-optic attenuation while the latter is limited by atmospheric extinction and diffraction. Here, we propose a hybrid network and protocol that outperform both ground- and satellite-based designs and lead to high-fidelity entanglement at a continental or even global scale.
\end{abstract}

\maketitle

\noindent
\section{I. Introduction}
The ongoing development of quantum technologies calls for advances in quantum communication. 
The basis of quantum communication is entanglement, which, once established, can allow the transmission of quantum information over arbitrary distances \cite{gisin2007quantum}. 
Multiple small-scale quantum networks have been successfully implemented  \cite{chung2021illinois, bersin2024development, martin2023madqci, garcia2024strategies, chen2021integrated}. 
The prospect of developing large-scale quantum networks 
is motivated by applications ranging from quantum key distribution \cite{xu2020secure, cao2022evolution} and distributed quantum computing \cite{cacciapuoti2020quantum} to  quantum navigation and sensing \cite{sanchez-burillo2012quantum, pirandola2018advances}. 
However, scalability remains a challenge due to difficulties associated with the generation of entanglement between distant parties, and thus
new approaches are needed to lay the groundwork for the future development of a quantum internet \cite{kimble2008quantum, wehner2018quantum}.

The most direct approach for distributing entanglement in a quantum network is by using optical fibers to transmit entangled photons \cite{wengerowsky2019entanglement,neumann2022continuous}. Due to the exponential loss of photons with distance, establishing long-distance entanglement requires intermediate quantum repeaters \cite{briegel1998quantum,azuma2023quantum}. Such devices,  which are currently under development,
are able to 
generate entangled states,
transmit/receive entangled photons,
perform quantum operations, and/or store qubits.
Within this approach, entanglement is first directly distributed over short distances.
Long-distance entanglement is then established through entanglement swapping performed at intermediate quantum repeaters. However, existing quantum repeater technology currently limits entanglement distribution over long distances \cite{avis2023requirements}. 

An alternative approach is to use satellites as sources of entangled photon pairs that are sent to two distant detection stations on the ground \cite{vallone2015experimental}. Entanglement distribution is successful if both ground stations detect photons of the same pair \cite{yin2017satellite}. The challenge with this approach is that photons are subject to diffraction loss, where the survival probability scales with the inverse square of the distance, and absorption and scattering in Earth's atmosphere.
In 2016, China launched Micius, the first quantum communication low Earth orbit (LEO) satellite \cite{lu2022micius}, which was
equipped with an entangled-photon-pair source and operated in an orbit 500 km from the ground. Entanglement distribution was established between two ground stations 1200 km apart from each other \cite{yin2017satellite}. Moreover, this LEO satellite was successfully used to 
enable quantum key distribution between China and Europe that are 7600 km apart on Earth \cite{liao2018satellite}. 
Subsequent work has also developed scheduling policies for entanglement distribution using groups of LEO satellites \cite{williams2024scalable}.

While using LEO satellites to distribute entanglement results in low photon loss, certain challenges with this technology currently impact its scalability 
\cite{lu2022micius}. First, LEO satellites travel at high speeds relative to the ground, leaving  only short time windows for interaction with the corresponding stations, and thus requiring high-precision acquisition, pointing, and tracking systems.
Second, the short distances between LEO satellites and the ground limit the area that can be covered by a single satellite, making it impossible for a satellite to establish entanglement between stations that are far apart. 
This second shortcoming can be addressed by allowing the satellite to temporarily store qubits as a trusted node, but this could compromise security \cite{wengerowsky2018entanglement}. An alternative to mitigate both shortcomings is to use geostationary (GEO) satellites, positioned at 36\,000 km above the equator in a fixed position relative to the ground \cite{miao2007feasibility}. A single GEO satellite
can cover more than one-third of the Earth's surface, but the loss caused by diffraction is $10^4$ times higher than for Micius,  which limits the entanglement distribution rate.  

In this article, we investigate a hybrid quantum network consisting of a combination of optical fibers and {\it medium} Earth orbit (MEO) satellites. The goal is to benefit from the short-distance advantage of optical fibers and the long-distance advantage of satellites. Our focus on MEO satellites---chosen to be 10\,000 km above ground---seeks to strike a compromise between low photon loss and high spatio-temporal coverage. Using the contiguous United States as a model system, we show that the hybrid protocol provides better large-scale quantum communication performance than both optical-fiber-based and satellite-based protocols.

\begin{figure}[t]
    \centering
    {\includegraphics[width=0.99\linewidth]{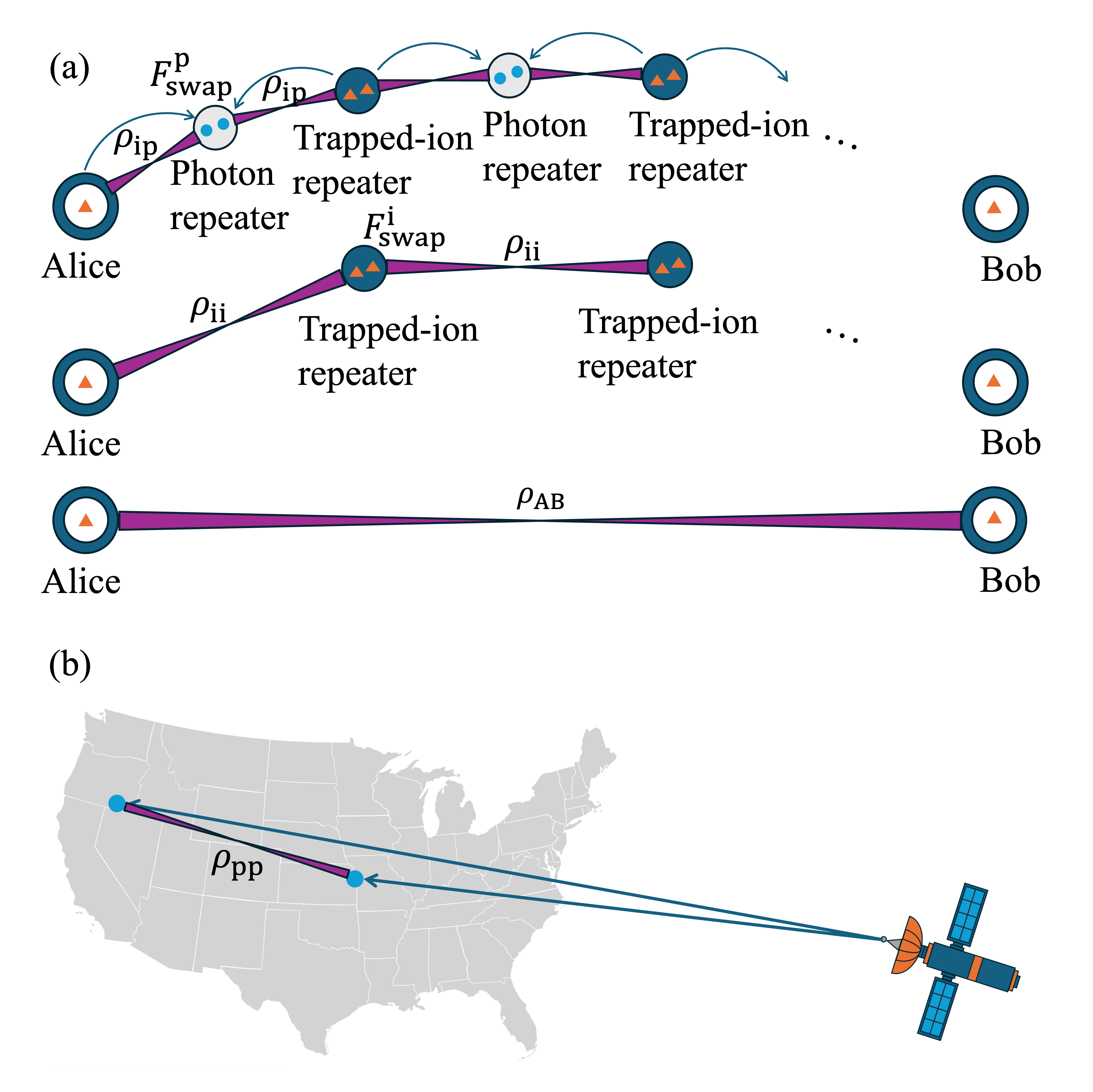}
    \caption{Idealized diagram of ground-based and satellite-based entanglement distribution. 
    (a) Optical-fiber path between Alice and Bob equipped with evenly distributed trapped-ion repeaters and intermediate photon repeaters. Top to bottom: end-point entanglement is achieved by ion-entangled photon transmission followed by entanglement swapping at the photon repeaters and then at the trapped-ion repeaters. (b) Satellite-mediated entanglement generation, where entanglement between ground stations is established by detecting entangled photon pairs emitted by the satellite.}
    \label{fig:schematic}}
\end{figure}

\noindent
\section{II. Optical-fiber transmission}
Because the number of photons successfully transmitted through an optical fiber decreases exponentially with distance, various quantum repeater designs have been proposed to compensate for this loss \cite{azuma2023quantum}. In particular,
a promising trapped-ion quantum repeater has been recently demonstrated in an entanglement distribution experiment 
between points separated by 50 km of optical fiber \cite{vrutyanskiy2023telecom}. A trapped-ion repeater includes at least two trapped ions and the ability to emit photons entangled with each ion. 
The trapped ions in the repeaters serve as quantum memories storing qubits.
Entanglement swapping between the ions can then be implemented through deterministic Bell-state measurement (DBSM). 

The ground portion of our hybrid quantum network will consist of a natural extension of this scheme, in which long-distance entanglement distribution is achieved using {\it multiple} quantum repeaters  [Fig.~\ref{fig:schematic}(a)].
In this extended scheme, a photon repeater is placed between every pair of trapped-ion repeaters (including the end nodes) \cite{vrutyanskiy2023telecom}. The photon repeaters can detect photons and perform quantum entanglement swapping using probabilistic Bell-state analysis. Alice and Bob (the end nodes) are assumed to have one trapped ion and, as trapped-ion repeaters, the ability to emit ion-entangled photons. 
Thus, the system consists of two end nodes,
$n_\text{p}$ photon repeaters, and
$n_\text{i}=n_\text{p}-1$ trapped-ion repeaters for $n_\text{p}\ge 1$ ($n_\text{i}=0$ otherwise). In this system, ion-photon entangled states $\rho_{\text{ip}}$ are converted into ion-ion entangled states $\rho_{\text{ii}}$ and then to an end-to-end entangled state $\rho_{\text{AB}}$ shared by Alice and Bob.

\begin{figure*}[t]
    \centering
    {\includegraphics[width=0.8\linewidth]{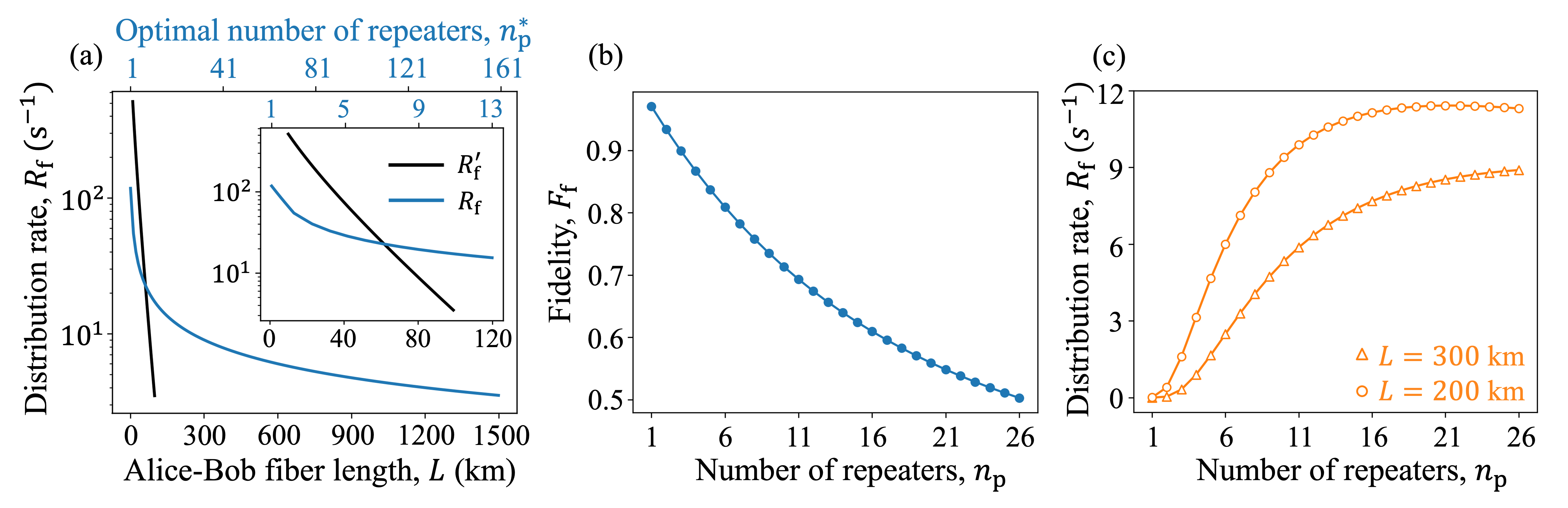}
    \caption{Entanglement distribution rate and fidelity through optical fibers.
    (a) Distribution rate without repeaters (black) and for an optimal number of evenly placed photon (and thus intermediate trapped-ion) repeaters (blue) as a function of the optical fiber length $L$. The optimal number of photon repeaters is marked on the top axis. 
    Inset: magnification of both curves for small $L$. 
    (b) Resulting end-to-end fidelity as the number of evenly placed repeaters is varied. (c) Corresponding entanglement distribution rate for two fixed values of $L$ as the number of repeaters is varied.}
    \label{fig:fiber_swapping}}
\end{figure*}

To proceed, we temporarily assume the ideal case in which the repeaters are evenly separated along the optical fiber path, as shown in Fig.~\ref{fig:schematic}(a). In this section, we also assume that each trapped-ion repeater has only two trapped ions (generalization to an arbitrary number is considered below). For $n_\text{p}\ge1$, the average time needed to establish one entanglement between Alice and Bob
is \cite{sangouard2009quantum}
\begin{equation}
	T_\text{f}(L,n_\text{p}) = \left(\frac{L}{n_{\text{p}} c}+\tau\right) \frac{3^{\nu}}{2^{\nu-1}P^2 \eta^2(L)}, 
\label{eq:ion_repeater}
\end{equation}
where $L$ is the fiber length between the end nodes, and the other symbols are defined as follows. The terms in parentheses account for the time $\tau$ needed to emit an entangled photon and the time to transmit the photon to an adjacent photon repeater, where $c=2\times 10^8$ m/s is the speed of light in the optical fiber; note that a factor of 1/2 is canceled because the trapped-ion nodes wait to hear  back from photon repeaters whether the photon was detected or not before attempting to send a new photon. The term $\frac{1}{2}P^2 \eta^2$ is the probability that 
both photons concurrently directed to a photon repeater
are successfully received, detected, and projected into a Bell state; here, $\eta(L) = 10^{-\gamma L/2n_{\text{p}}}$ is the fiber transmission rate for an attenuation rate $\gamma$, $P$ is the photon-detector efficiency, and 1/2 is the probability of a successful photon Bell-state measurement. 
The waiting time to establish a successful entanglement between {\it one} pair of adjacent trapped-ion nodes follows a geometric distribution; thus, the additional factor $(3/2)^\nu$ is a good approximation for the waiting time before {\it all} pairs in the path are ready to be entangled, where $\nu=\log_2{n_{\text{p}}}$.
The time for local quantum operations associated with entanglement swapping is assumed to be negligible, with entanglement swapping having a success rate of 1/2 for photon repeaters (as mentioned) and 1 for trapped-ion repeaters.
The average entanglement distribution rate is then given by $R_\text{f}=1/T_\text{f}$. 
As a reference, we note that 
the average time required to establish an entanglement between Alice to Bob without any intermediate repeater~is
\begin{equation}
    T_\text{f}'(L,0) = \left(\frac{2L}{c}+\tau\right) \frac{1}{P \eta'(L)}, 
\end{equation}
where $\eta'(L)=10^{-\gamma L}$, and the average entanglement distribution rate is thus given by $R_\text{f}'=1/T_\text{f}'$.

Figure~\ref{fig:fiber_swapping}(a) shows the resulting distribution rate $R_\text{f}$ for the optimal number of repeaters $n_{\text{p}}^*$ as the end-to-end fiber length $L$ is varied, obtained by minimizing $T_{\text{f}}$ as a function of $n_{\text{p}}$ in Eq.~(\ref{eq:ion_repeater}). We also show the distribution rate $R_\text{f}'$, which is higher for distances smaller than $61.7$ km (inset). Note that at this distance, the optimal number of photon repeaters transitions from 0 to 7 (rather than to 1); this gap comes from the competition (as $n_\text{p}$ is increased) between the increase in the entanglement distribution rate
with neighbors and the decrease in the probability of successful Bell-state measurements for all photon repeaters in the path.  
Following \cite{vrutyanskiy2023telecom}, our simulations assume that photons are emitted entangled with $^{40}$Ca$^+$ ions and converted to 1550 nm,
which corresponds to the lowest attenuation in the optical fiber, $\gamma=0.0173$ km$^{-1}$. The other parameters are $P=0.21$ and $\tau=175$ $\mu$s, where the efficiency of photon conversion is incorporated into $P$. At large distances, repeaters allow the decay of the entanglement distribution rate to be substantially slower than the exponential one observed without repeaters, but the optimal number of repeaters increases linearly with distance. Note that for a fixed fiber length, the entanglement distribution rate increases monotonically as $n_{\text{p}}$ increases before reaching the optimum at $n_{\text{p}}^*$ [Fig.~\ref{fig:fiber_swapping}(c)].

A competing factor to be accounted for is the end-to-end fidelity as the number of repeaters increases. We follow the model discussed in \cite{meraner2020indistinguishable} to calculate the fidelity as a function of $n_{\text{p}}$. We assume that the initially generated ion-photon entanglement is a mixed state of the form $\rho_{\text{ip}} = F_0 |\Psi^-\rangle\langle \Psi^-| + (1-F_0) |\Psi^+\rangle\langle \Psi^+|$, where $|\Psi^{\pm}\rangle$ are Bell states and $F_0$ is the pairwise fidelity [Fig. \ref{fig:schematic}(a)]. A successful Bell-state measurement on a photon pair, which occurs with a probability $1/2$, leads to a new mixed entangled state $\rho_{\text{ii}} = F_{\text{ii}} |\Psi^-\rangle\langle \Psi^-| + (1-F_{\text{ii}}) |\Psi^+\rangle\langle \Psi^+|$ shared by two neighboring trapped-ion nodes. The ion-ion entanglement fidelity is $F_{\text{ii}}=[1+V(1-2F_0)^2]/2$, where $V=2F_{\text{swap}}^{\text{p}}-1$ is the photon interference-visibility. Each intermediate trapped-ion repeater node can then perform a DBSM on the two trapped-ion qubits they maintain. Given $\rho_{\text{i}_1\text{i}_2}$ and $\rho_{\text{i}_3\text{i}_4}$, an ideal DBSM will give $\rho_{\text{i}_1\text{i}_4}$ by projecting trapped-ions 2 and 3 onto a Bell state and tracing them out. If we account for the depolarizing noise caused by the DBSM, the resulting quantum state will be $\tilde{\rho}_{\text{i}_1\text{i}_4}(\rho_{\text{i}_1\text{i}_4},F^{\text{i}}_{\text{swap}})$, where $F_{\text{swap}}^{\text{i}}$ is the trapped-ion entanglement-swapping fidelity and 
\begin{equation}
	\tilde{\rho}(\rho,F)= F\rho + \frac{1-F}{3}(S_z\rho S_z + S_y\rho S_y + S_x\rho S_x). 
\end{equation} 
Here, $S_j\in [I \otimes\sigma_x, I \otimes\sigma_y, I \otimes\sigma_z]$, where $\sigma_j$ are the Pauli matrices and $I$ is the identity matrix. 
The fidelity of the final end-to-end entangled state is obtained after completing all the entanglement-swapping operations on the intermediate trapped-ion nodes. 
Through the entire process, the fidelity of the established entanglement can decay due to the decoherence of qubits stored in the quantum memories. For an arbitrary initial state $\rho_0$, this can be modeled by considering dephasing noise of the form 
\begin{equation}
    {\rho}(t) = q(t) \rho_0 + [1-q(t)]S_z^\dagger \rho_0 S_z, 
\end{equation}
where $q(t)=(1-e^{-t^2/\tau_q^2})/2$ and $\tau_q$ is the decoherence time. A recent study showed that the decoherence time of trapped-ion entanglement can be longer than 10 s \cite{drmota2023robust}, which is much longer than the relevant time needed to establish one end-to-end entanglement discussed in this study. Therefore, we assume the impact of decoherence to be negligible in our calculations, leading to no fidelity loss. An analysis of the impact of non-negligible decoherence is included in Appendix D. 

The end-to-end entanglement fidelity $F_\text{f}$ decreases as the number of quantum repeaters increases. Entanglement is completely lost for $n_{\text{p}}>26$ when the three parameters are $F_0$ = $F_{\text{swap}}^{\text{p}}$ = $F_{\text{swap}}^{\text{i}}$ = 0.99 (as assumed throughout unless otherwise noted) [Fig.~\ref{fig:fiber_swapping}(b), blue] \cite{noteSM1}. 
Therefore, entanglement distillation is necessary to achieve high end-to-end fidelity, which we implement using the recurrence method for mixed-state distillation. Given two entangled mixed states with fidelities $F_1$ and $F_2$ shared by the same pair of nodes, a successful purification yields a single entanglement with a higher fidelity $F_{\text{p}}=(10F_1F_2-F1-F2+1)/(8F_1F_2-2F_1-2F_2+5)$. The probability of completing such a 2$\rightarrow$1 purification successfully is $(8F_1F_2-2F_1-2F_2+5)/9$. With multiple copies of entanglement shared by two neighboring nodes, the 2$\rightarrow$1 purification can be iteratively repeated until only one entangled state remains for the given pair of nodes. The distillation process will allow us to reduce the optimal number of repeaters used in Fig.~\ref{fig:fiber_swapping}, which was generated without considering purification. 

\begin{figure*}[t]
    \centering
    \includegraphics[width=0.9\textwidth]{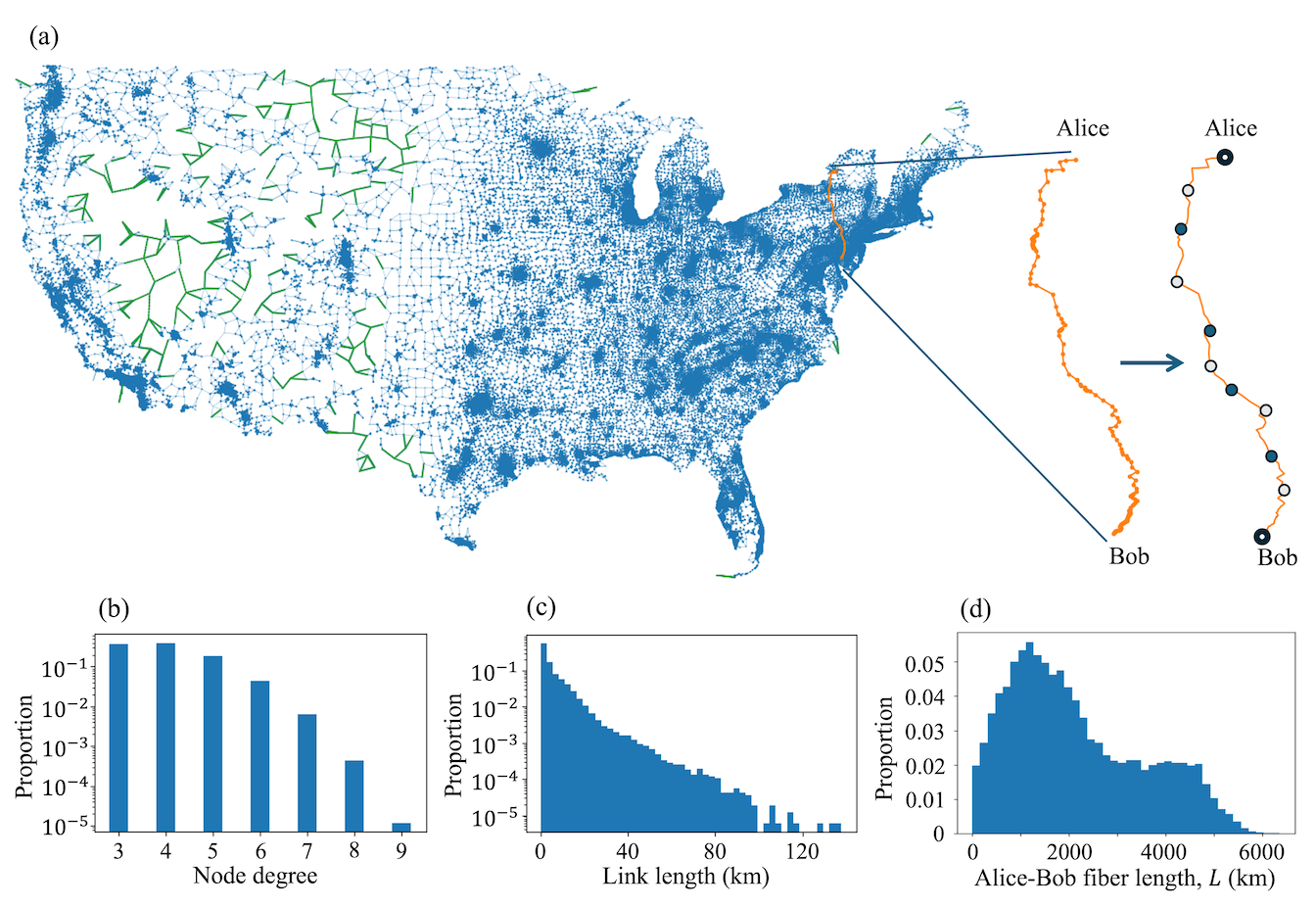}
    \caption{Model optical-fiber network of the contiguous U.S. (a) Network structure, where
    each node represents a census tract and each link represents an optical fiber. Links that require intermediate repeaters, namely those longer than 61.7 km, are marked green. The orange path illustrates an entanglement distribution route between end nodes Alice and Bob. (b-d) Histograms of node degrees (b), link lengths (c), and optical-fiber path lengths between $10\,000$ randomly sampled end-node pairs (d) in the network. } 
    \label{fig:fiber_map}
\end{figure*}

\noindent
\section{III. Satellite transmission}
We consider an MEO satellite orbiting 10\,000 km above ground. The satellite is equipped with an entangled-photon source that can generate $N_{\text{s}}$ entangled photon pairs per second. The photon pairs are split into two collimated beams and are then sent toward two ground stations as shown in Fig.~\ref{fig:schematic}(b). An entanglement is successfully established between two ground stations if both stations detect photons from the same entangled pair. The ground stations will then store the quantum states of the received photons in single-photon quantum memories [which also use
trapped ions, but are different from the trapped-ion memories that operate by emitting photons in Fig.~\ref{fig:schematic}(a)] \cite{wang2019efficient}.
Free-space diffraction and atmospheric extinction (absorption and scattering) are accounted for as the main factors causing losses during transmission \cite{pirandola2021satellite}. Other losses come from limitations in the transmitting and receiving telescopes.

\begin{figure*}[t]
    \centering
    \includegraphics[width=0.9\linewidth]{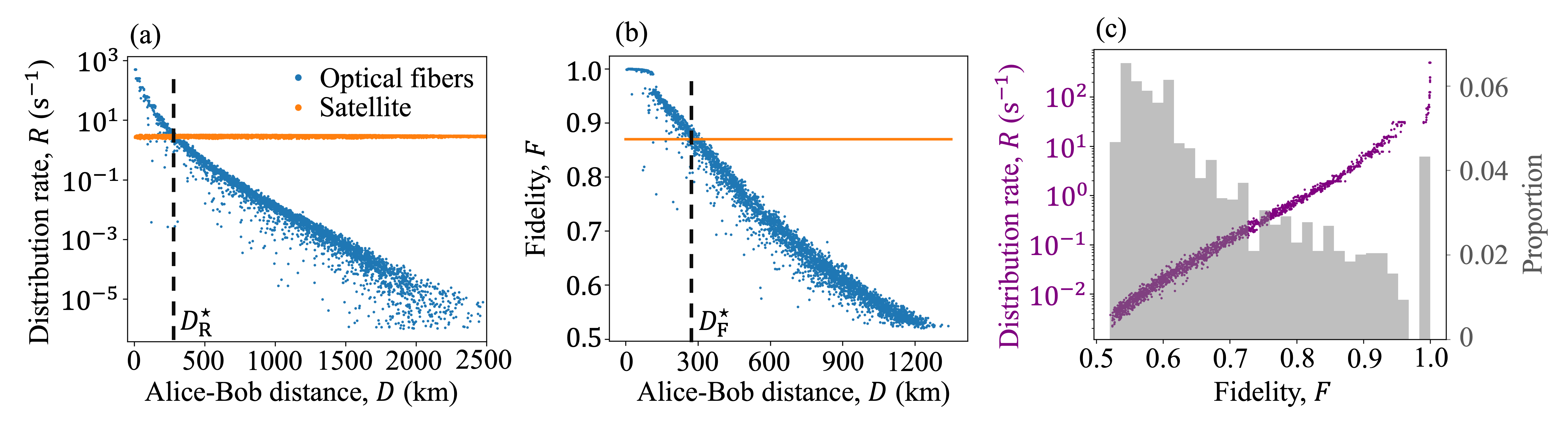}
    \caption{Entanglement distribution efficiency through the exclusive use of optical fibers or MEO satellites for 10\,000 randomly sampled Alice-and-Bob pairs in the network shown in Fig.~\ref{fig:fiber_map}(a). (a) Distribution rate through optical fibers (blue) and a satellite (orange).
    For a fair comparison, the fidelity for the satellite and target fidelity for the ground network are both assumed to be $0.87$. 
    (b) Fidelity of the end-to-end entanglement for the time needed to establish one entanglement via satellite, color-coded as in (a). The points $D^\star_\text{R}$ and $D^\star_\text{F}$ mark the distances where the relative efficiency inverts. Repeaters are placed to achieve optimal distribution rate in (a) and optimal fidelity in (b). 
    (c) Scattered plot (purple) and histogram (grey) of the distribution rate in (a) and fidelity in (b). 
    }
    \label{fig:comparison}
\end{figure*}

Diffraction broadens the beam waist during transmission. The spot size of the beam after traveling a distance $z$ can be calculated as 
\begin{equation}
	w_{\text{d}}(z) = w_0\sqrt{(1+z^2/z_{\text R}^2)}, 
\end{equation}
where $z_{\text R}=\pi w_0^2\lambda^{-1}$ is the Rayleigh range, $\lambda$ is the wavelength, and $w_0$ is the initial beam width. 
The slant range $z$ can be easily calculated given the longitudes and latitudes of the ground station and the subsatellite point (the projection point of the satellite on the Earth's surface) by converting them to Cartesian coordinates. 
Given a telescope receiver of a radius $a_{\text{r}}$, the fraction of the initial beam that can be detected is 
\begin{equation}
	\eta_\text{d}(z)=1-e^{-2a_{\text{r}}^2/w_{\text{d}}^2}, 
\end{equation}
which can be approximated as 
\begin{equation}
	\eta_{\text{d}}=\frac{2a_{\text{r}}^2}{w_{\text{d}}^2}
\end{equation}
for the far field $z\gg z_{\text R}$ relevant for MEO satellites.
Absorption and Rayleigh/Mie scattering occur when the beam travels through the atmosphere (mainly the $30$ km layer above the ground). The fraction of the beam surviving absorption and scattering is
\begin{equation}
	\eta_{\text{atm}}(\theta)\approx \big[\eta_{\text{atm}}^{\text{zen}}(\infty)\big]^{\text{sec}\,\theta}, 
\end{equation}
for a zenith angle $\theta\lesssim 1$ rad  and $\eta_{\text{atm}}^{\text{zen}}(\infty)\approx 0.967$.

The resulting satellite entanglement distribution rate is then 
\begin{equation}
	R_{\text{s}} = N_{\text{s}} \eta(z_1, \theta_1) \eta(z_2, \theta_2), 
\end{equation}
where $\eta(z, \theta)$ is the photon transmission rate between the satellite and 
one ground station. This rate accounts for the efficiency $\eta_{\text{tran}}$ of the transmitting telescope and the optics efficiency $\eta_{\text{rec}}$, detection efficiency $\eta_{\text{dec}}$, and spectral-filter efficiency $\eta_{\text{filt}}$ of the receiving telescope at the ground station \cite{gruneisen2017modeling,gruneisen2021adaptive}. The photons also need to be converted to a wavelength and spatiotemporal mode compatible with the quantum memories at the ground stations. Since the conversion efficiency can vary depending on the type of memory used, we assume a perfect conversion for simplicity; this factor can be accounted for in future studies.
Combining all the terms, the photon transmission rate reads $\eta(z, \theta)=\eta_{\text{tran}}\eta_{\text{d}}(z)\eta_{\text{atm}}(\theta) \eta_{\text{rec}} \eta_{\text{det}} \eta_{\text{filt}} \eta_{\text{coup}}$, where the extra $\eta_{\text{coup}}$ is the single-mode optical-fiber coupling efficiency enhanced by a high-bandwidth adaptive-optics system \cite{cameron2024adaptive,scarfe2025fast}. 

The photon source used on Micius operated with a rate of $5.9$ million entangled photon pairs per second and a fidelity of $0.907\pm 0.007$. The successfully detected photon pairs were still entangled with a fidelity of $0.869\pm 0.085$. Since noise from the environment is negligible when entangled photons are transmitted through the vacuum (before reaching the Earth's atmosphere), an MEO satellite with the same source would distribute entanglement with the same fidelity.
Due to the much larger serving area of MEO satellite, we assume the satellite is equipped with multiple such entangled photon sources, resulting in a total generation rate of $N_{\text{s}}=6\times 10^7$ with a fidelity similar to Micius, namely $F_{\text{s}}=0.87$.

\noindent
\section{IV. Hybrid satellite and optical-fiber network}
To study the hybrid protocol for entanglement distribution combining both optical fibers and satellites in a large-scale quantum network, we first focus on the physical aspects of a representative case. We model a hybrid quantum network for the contiguous United States using MEO satellites and a ground optical-fiber network designed using the
population-distribution data from the 2020 national census \cite{census2020}. 
The ground network has a total of $83\,047$ nodes, each representing the latitude and longitude coordinates of the centroid of a census tract region (or the point within the region that is closest to the centroid). A census tract is a subdivision of a county that is used by the U.S. Census Bureau for organizing and analyzing census data; each includes an average of 3\,957 residents.

Figure~\ref{fig:fiber_map} shows the resulting network, in which each node is assumed to be connected by two-way optical fibers (undirected links) to its three closest neighbors. 
To allow at least one path of optical-fiber links connecting any two nodes (i.e., to form a single network connected component),
an extra 24 links are added while minimizing the total fiber length [Fig.~\ref{fig:fiber_map}(a)],  leading to a total of $162\,860$ links.
The number of neighbors connected to a node (i.e., the node degree) ranges from 3 to 9, with an average of $3.9$ [Fig.~\ref{fig:fiber_map}(b)].

The length of each link is approximated by the Euclidean distance between the end nodes. 
The distribution of link lengths shows that
$99.84\%$ of all the links are shorter than $61.7$ km [Fig.~\ref{fig:fiber_map}(c)]. This distance corresponds to the intersection in Fig.~\ref{fig:fiber_swapping}(a) and thus,
to achieve optimal distribution rates, intermediate repeaters are added to longer links [marked green in Fig.~\ref{fig:fiber_map}(a)].  
The distribution of 
shortest fiber-length paths connecting
randomly selected Alice-and-Bob end-node pairs shows that the vast majority of pairs are separated by more than $1\,000$ km [Fig.~\ref{fig:fiber_map}(d)].

All nodes (including the intermediate repeaters on long links) can play the role of a trapped-ion repeater or a photon repeater, depending on the Alice-Bob pair establishing end-to-end entanglement. They are equipped with trapped ions, can emit and receive photons, and can perform entanglement swapping on photons and trapped ions. 
Furthermore, we assume that the trapped-ion repeaters can benefit from having multiple pairs of trapped ions, which allows for multiplexing \cite{azuma2023quantum}.
In our simulations below, each trapped-ion repeater node is considered to have $20$ trapped ions ($10$ for each direction of entanglement along the given path, as opposed to $1$ used thus far), which helps boost the efficiency of entanglement distribution.  

The communication between end nodes in the ground network is assumed to take place through the optical-fiber path with the shortest length. Along this path, a greedy algorithm is applied to find the {\it optimal  placement} of both photon and trapped-ion repeaters (i.e., determining which repeaters should be active or bypassed as well as the role to be played by the active repeaters).
In this optimal placement, we maximize the end-to-end fidelity or the entanglement distribution rate, depending on the question under consideration, under the constraint of only using repeaters the network is instrumented with
(the optimal placement will thus generally be unevenly distributed along the shortest path between the end nodes).
Figure~\ref{fig:fiber_map}(a) highlights the shortest fiber path for a choice of Alice and Bob, which has $547.5$ km and $132$ intermediate nodes, where the optimization of the end-to-end fidelity leads to the optimal placement of only 5 photon repeaters (alternated by 4 trapped-ion repeaters).

\begin{figure*}[t]
    \centering
\includegraphics[width=0.9\linewidth]{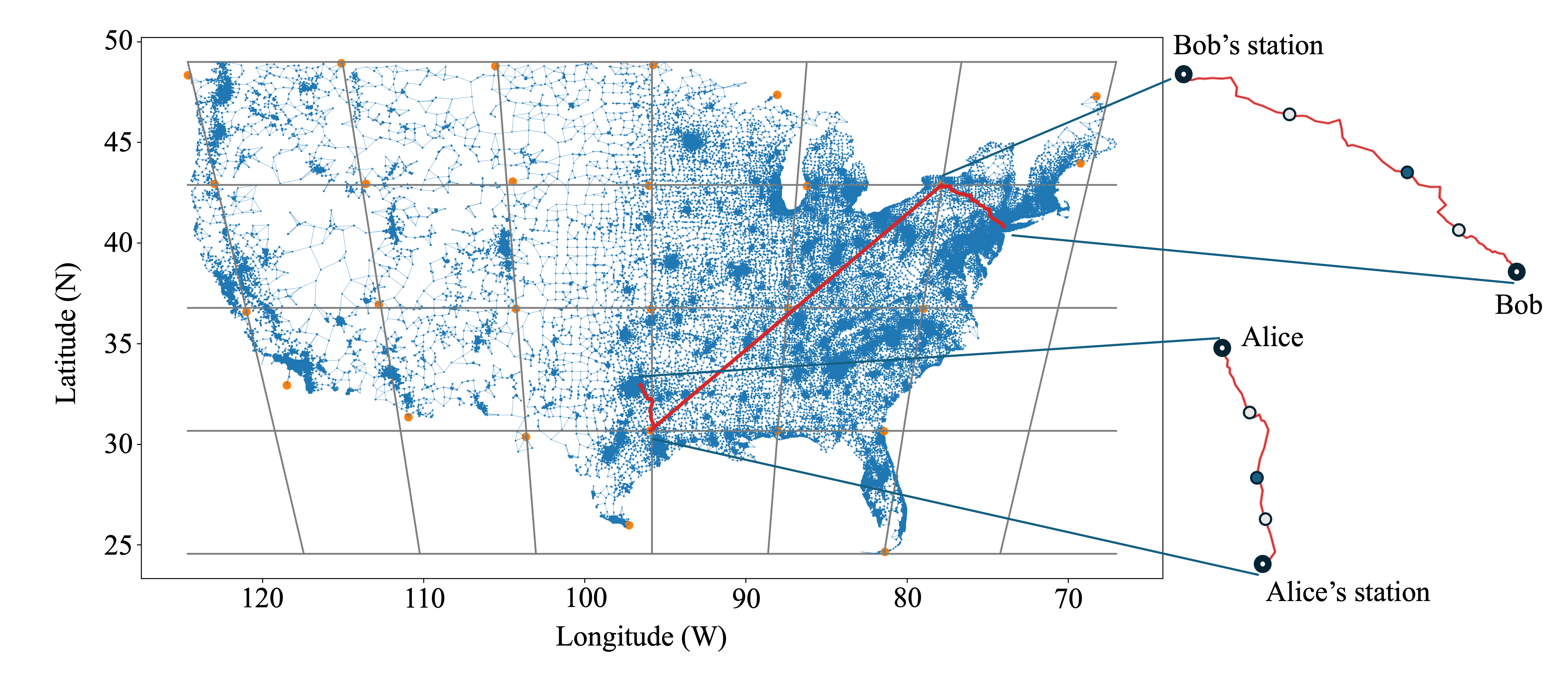}
    \caption{Hybrid network protocol for sparse placement of satellite ground stations. The orange dots indicate ground stations placed on nodes in the network closest to the intersection points of a given grid. In the hybrid protocol, entanglement requests between distant Alice and Bob nodes are first routed (through shortest paths) to their nearest ground stations; entanglement is then established using a combination of satellite links and optical-fiber links.}
    \label{fig:satellite_stations}
\end{figure*}

To devise a hybrid network, we now compare the exclusive use of optical fibers just discussed with the exclusive use of MEO satellites. 
In the latter case, and only for the purpose of this comparison, each end node would have to be equipped with a ground station (a telescope) able to detect entangled photons. The satellite is then assumed to generate entangled photon pairs and send them directly to the ground stations of Alice and Bob. 
The photon source on the satellite is considered to generate $N_{\text{s}}=6\times 10^7$ entangled down-converted photon pairs per second, where the wavelength is $\lambda=810$ nm and the beam width is $w_0=0.15$ m. The radius of the satellite ground station telescopes is taken to be $a_\text{r}=0.5$ m (see Appendix C for the performance improvement enabled by larger $a_\text{r}$). 
The other parameters are taken as $\eta_{\text{tran}}$ = $\eta_{\text{rec}}$ = $0.5$, $\eta_{\text{filt}}$ = $\eta_{\text{dec}}$ = 0.9, and $\eta_{\text{coup}}$ = 0.7.

Figure \ref{fig:comparison}(a) compares the entanglement distribution rates using only optical fibers or only MEO satellites, for the parameters above in each case. To facilitate comparison,
we consider the rate as a function of the geographical distance $D$ (rather than fiber length) between the two end nodes. The satellite entanglement distribution rate $R_\text{s}$ is almost constant over $D$, at approximately 3 s$^{-1}$. While using optical fibers, entanglement distillation is performed for a target end-to-end entanglement with a fidelity equal to the satellite fidelity, $F_{\text{s}}$. 
The distribution rate decreases as the end-to-end geographical distance $D$ increases and falls below the satellite rate when $D>D^\star_\text{R}\approx 280$ km. 
On the other hand, Fig.~\ref{fig:comparison}(b) shows the fidelity of the end-to-end entanglement via fibers $F_\text{f}$ when the time for establishing one copy of entanglement between Alice and Bob is fixed (to be the average time needed to establish one entanglement pair via satellite).
The end-to-end fidelity via fibers decreases rapidly as $D$ increases, and it drops below $F_\text{s}$ for $D>D^\star_\text{F}\approx 280$ km
and below $0.5$ for $D>1200$ km.   
Figure~\ref{fig:comparison}(c) summarizes the relationship between the distribution rate (for a given target fidelity) and the fidelity (for a fixed time) for all Alice-Bob pairs in Fig.~\ref{fig:comparison}(a)-(b). The distribution rate increases monotonically with fidelity, as both are strongly correlated with the distance $D$ between the Alice and Bob nodes.

\begin{figure*}
    \centering
    \includegraphics[width=\linewidth]{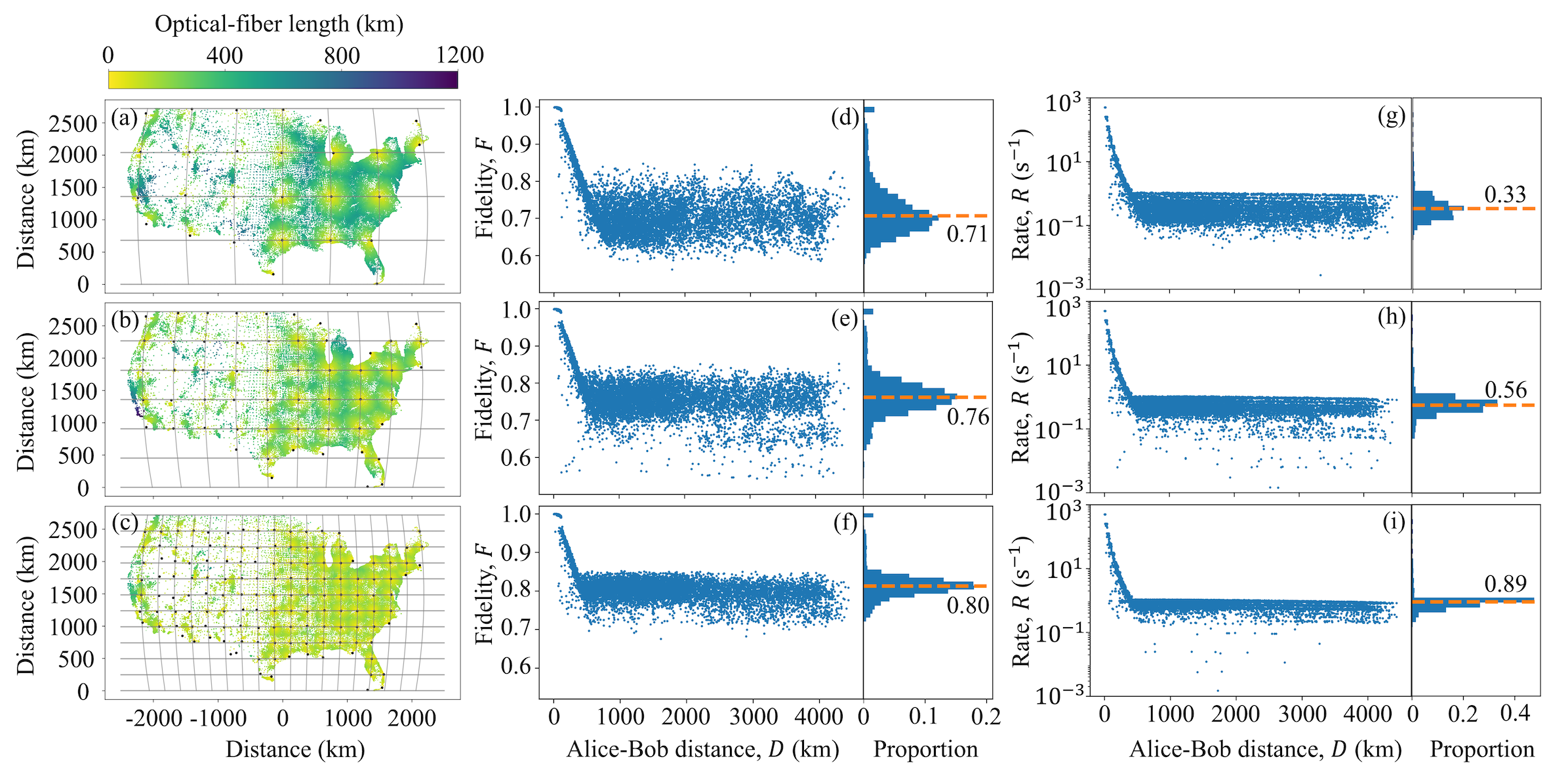}
    \caption{Performance of the hybrid network protocol.
    Three grid sizes are considered for the placement of satellite ground stations: $d= 700$ km (top row), $d=450$ km (middle row), and $d=240$ km (bottom row). (a)-(c) Map of the ground network, in which the nodes are color coded according to their distance to the closest satellite ground station (black dots). Distances are marked relative to the point with a latitude of 24.6$^\text{o}$ N and a longitude of 95.8$^\text{o}$ W. (d)-(f)
    Scatter plot (left) and histogram (right) of the
    end-to-end fidelities for the hybrid protocol. (g)-(i)
    Scatter plot (left) and histogram (right) of the
    entanglement distribution rates for the hybrid protocol for a target fidelity of 0.87. In (d-i), the repeaters are optimally placed as in Fig.~\ref{fig:comparison}. In all histograms, the dashed lines indicate the median. 
    }
\label{fig:satellite_stations_different_size}
\end{figure*}

While the MEO satellites are in orbits of 5.8 h around the globe, in our model we assume that at least one satellite has a subsatellite point within the approximate boundaries of the contiguous U.S. in Fig.~\ref{fig:fiber_map}(a) at each time. 
In Appendix E, 
we show that the satellite's location has only a small effect on the entanglement distribution rate and end-to-end entanglement fidelity. Therefore, we approximate its location as the middle point of the map (latitude 36.8$^\text{o}$ N, longitude 95.8$^\text{o}$ W) in our calculations.
Crucially, due to the reduced diffraction, the entanglement distribution rate for MEO satellites is substantially higher than for GEO satellites (for details, see Appendix A).

\noindent
\section{V. Hybrid network protocol}
The results above clearly show that not all pairs of end nodes can communicate most efficiently through optical fibers only or satellites only: the ground network is more efficient at short distances while the satellites are superior at longer distances. However, the distances $D^\star_\text{R}$ and $D^\star_\text{F}$ at which the relative efficiency inverts  depend on the placement of the satellite ground stations. In Fig.~\ref{fig:comparison}, we considered the limit case in which every node was equipped with one such station, which offers a lower-bound estimation for $D^\star_\text{R}$ and $D^\star_\text{F}$. In a more realistic scenario, we can equip the network with a grid of stations, as shown in Fig.~\ref{fig:satellite_stations} 
for a grid spacing $d$ of approximately $700$ km (which limits the maximum distance between an end node and the nearest station to $\approx\! 495$ km).

We can now complete the hybrid network protocol. When two distant end nodes, Alice and Bob, request end-to-end entanglement, they will first find the closest satellite ground stations and establish entanglement through the fibers with those stations. Meanwhile, entanglement between the two ground stations equipped with single-photon quantum memories
is established using a satellite. 
 Following the recent demonstration of high efficiency and fidelity of single-photon quantum memories \cite{wang2019efficient}, below we assume no fidelity loss when the quantum states are stored. 
The final entanglement swappings at the two stations will then establish the desired end-to-end entanglement between Alice and Bob. The red path in Fig.~\ref{fig:satellite_stations} shows an example.

Figure \ref{fig:satellite_stations_different_size} shows a visual representation of these results for $d=700$, $450$, and $240$ km.
To facilitate interpretation, the map is now parameterized by geographic distance (rather than latitude and longitude), and the nodes are color-coded by the length of the shortest optical-fiber path to the closest satellite ground station  [Fig.~\ref{fig:satellite_stations_different_size}(a)-(c)].
For a fair comparison, the time required to establish entanglement via satellite is used as the time available for entanglement distribution and distillation (between endpoints or between endpoints and the closest satellite ground stations) via optical fibers. 
The fidelity decreases slowly as the number of satellite ground stations is reduced
[Fig.~\ref{fig:satellite_stations_different_size}(d)-(f)], with a median of $0.71$ for $d=700$ km; this should be contrasted with the fidelity of 0.87 for the idealized case of $d=0$ km in Fig.~\ref{fig:comparison}(b). 
On the other hand, the median of distribution rate decreases rapidly as $d$ is increased [Fig.~\ref{fig:satellite_stations_different_size}(g)-(i)]. This is the case because of the exponential loss of photons with the length of the path between end nodes mediated by the ground network, which in turn grows linearly with $d$.
In addition, the distribution rate varies substantially more widely as $d$ increases [note the logarithmic scale in Fig.~\ref{fig:satellite_stations_different_size}(g)-(i)].

To examine the impact of varying the spacing of the satellite ground stations, we first note that the distances
$D^\star_\text{R}$ and $D^\star_\text{F}$
are now defined by the intersection between optical fibers only and the combined use of optical fibers and satellite (as in the highlighted path in Fig.~\ref{fig:satellite_stations}), which are the inflection points in Fig.~\ref{fig:satellite_stations_different_size}(d)-(i).
Table \ref{table:grid} summarizes the trends in $D^\star_\text{R}$ and $D^\star_\text{F}$ as $d$ is varied. It follows that even when the entire contiguous U.S. is equipped with only $27$ stations, as for $d=700$ km in Fig.~\ref{fig:satellite_stations}, $D^\star_\text{F}$ and $D^\star_\text{R}$ remain approximately $800$ and $640$ km, meaning that the vast majority of node pairs would communicate through paths that involve both optical fibers and satellites.

\begin{table} [b]
\begin{center}
\caption{Distances $D_\text{F}^\star$ and $D_\text{R}^\star$ at which the relative efficiency inverts as grid spacing $d$ between ground stations (and thus the maximum distance $d/\sqrt{2}$ to closest satellite ground station) is varied.}
\label{table:grid}
\begin{tabular}{ r|r c c } 
\hline
$d$ (km) & $d/\sqrt{2}$ (km)
& $D_\text{F}^\star$ (km) & $D_\text{R}^\star$ (km)\\
\hline
\kern-1em $700$~ & ~ $495$ ~ & ~ $800$ ~ & ~ $640$ ~ \\ 
~ $450$ ~ & ~ $318$ ~ & ~ $580$ ~ & ~ $530$ ~ \\ 
~ $240$ ~ & ~ $140$ ~ & ~ $420$ ~ & ~ $420$
~ \\ 
~ $0$ ~ & ~ $0$ ~ & ~ $280$ ~ & ~ $280$ ~ \\
 \hline
\end{tabular}
\end{center}
\end{table}

\begin{figure}
    \centering
    \includegraphics[width=0.95\linewidth]{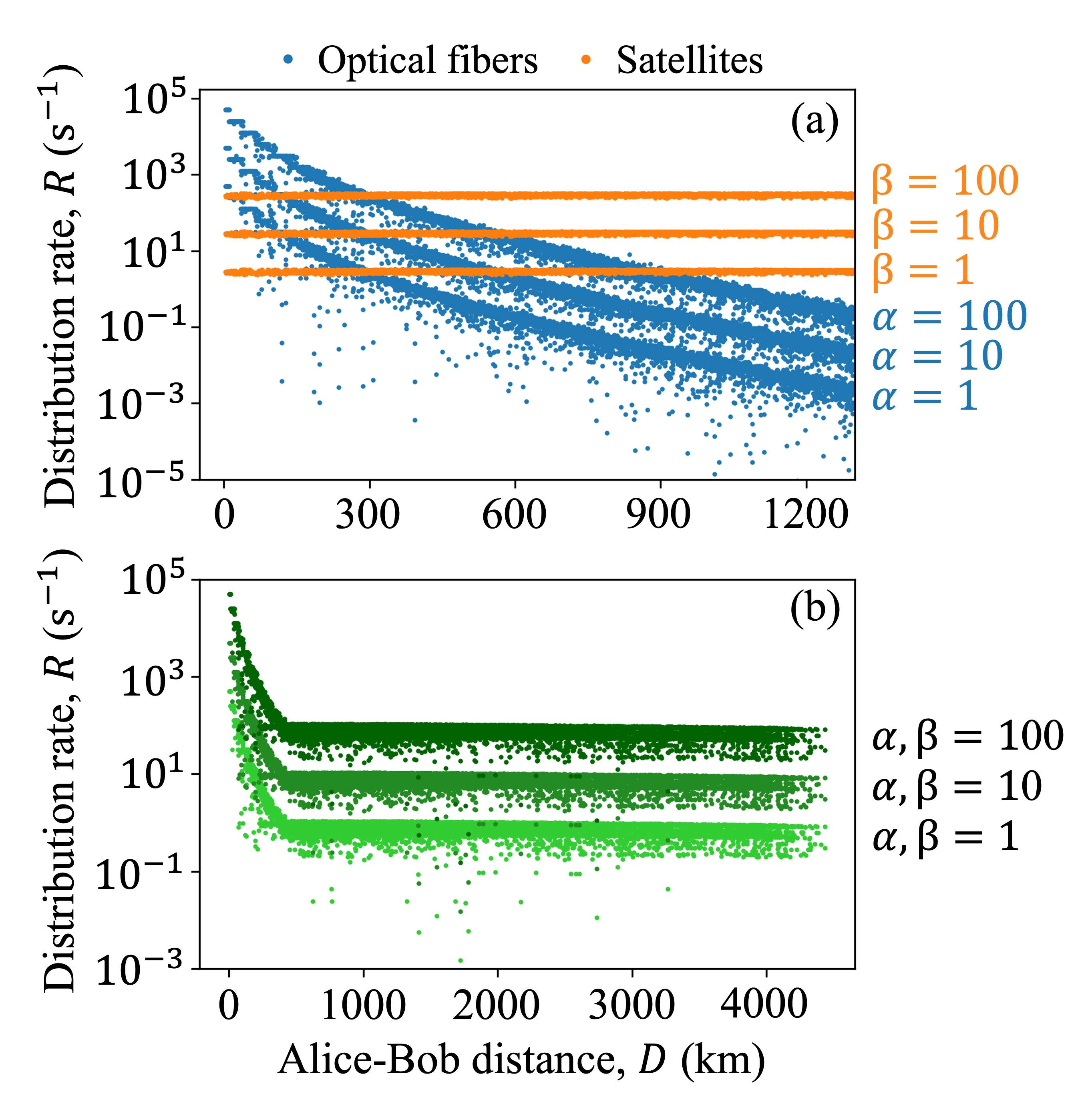}
    \caption{
    Improvement in entanglement distribution rate when the photon generation rates are increased.  
    (a) Counterpart of Fig.~\ref{fig:comparison}(a) 
    when the rates at the trapped-ion nodes and satellite are increased by factors $\alpha$ and $\beta$, respectively.
    (b) Resulting distribution rates for the hybrid protocol with $\alpha=\beta$ when the satellite ground stations are placed as in Fig.~\ref{fig:satellite_stations_different_size}(c). 
    }
\label{fig:alpha_beta}
\end{figure}

Finally, it is important to consider how the ability to generate and manipulate an increased number of entangled photon pairs, to be enabled by future technologies, would impact our results. 
Suppose the number of ions at each trapped-ion repeater is increased by a factor $\alpha$ and the photon emission rate $N_{\text{s}}$ at the satellite is increased by a factor $\beta$ relative to the rates considered in Figs.~\ref{fig:comparison} and \ref{fig:satellite_stations_different_size}.
It follows that the corresponding entanglement distribution rates will also increase by the same factor $\alpha$ or $\beta$, as shown in Fig.~\ref{fig:alpha_beta}(a), leading to the same intersections depicted in Fig.~\ref{fig:comparison}(a) when $\alpha=\beta$. Accordingly, when $\alpha=\beta$ in the hybrid protocol, the end-to-end fidelity in Fig.~\ref{fig:satellite_stations_different_size}(d)-(f) remains unchanged while the  entanglement distribution rate in Fig.~\ref{fig:satellite_stations_different_size}(g)-(i) increases by the same factor $\alpha$ [as illustrated in Fig.~\ref{fig:alpha_beta}(b)]. The latter is significant because the entanglement distribution rate is otherwise low, with an average of 6.3 s$^{-1}$ for the idealized case in Fig.~\ref{fig:comparison}(a). For $\alpha=\beta=100$, this number increases to the respectable figure of $630$ s$^{-1}$. 

\noindent
\section{VI. Conclusions}

The model for a hybrid quantum network conceived in this article is capable of establishing entanglement for nationwide quantum communication across the contiguous U.S. Our analysis demonstrates that a hybrid satellite-fiber design offers superior large-scale
communication 
performance compared to either an optical-fiber-based or a satellite-based design alone.
In our protocol, 
optical fibers are invoked for efficient entanglement distribution at short distances, while the satellites are required to maintain a nearly constant distribution rate over long distances. 

The hybrid model can be extended in several ways. 
First, our results are based on a uniform grid distribution of satellite ground stations. A strategic placement of the ground stations that accounts for population density would increase the average satellite-based entanglement distribution rate.
Second, our analysis considered only one satellite. Implementing instead an ensemble of satellites could reduce diffraction and atmospheric extinction and increase the rate of satellite-generated entanglements.
Third, while  we focused on medium Earth orbit satellites, 
performance can be further enhanced by considering a multilayer architecture in which satellites at different altitudes are used for communication between ground stations at different distances.
Fourth, we considered communication within the contiguous U.S., but similar hybrid network and protocol can be implemented at the global scale. The key step for a global design is to include satellite-to-satellite communication \cite{goswami2023satellite}, 
in which entangled photons can be transmitted through two or more satellites before reaching the ground. Altogether, these extensions would strengthen the overall results 
by increasing the entanglement generation rate, improving fidelity, and expanding the possible number of communicating parties.

The development of a practical large-scale quantum network is currently constrained by technological limitations.
At this stage, it is not yet clear what network architecture will prevail. We suggest that by focusing on the codesign of network and protocol, as considered here, one can identify the technologies that should be prioritized and the principles from classical communication networks that should be leveraged \cite{bacciottini2024leveraging}. Ultimately, this approach is relevant not only for the possible development of a quantum internet but also for the design of quantum sensing networks for high-precision physics.

\noindent
\section{Ackowledgements}
This work was supported by ARO MURI Grant No.~W911NF-21-1-0325. The views and opinions expressed herein are those of the authors and do not necessarily reflect those of the United States Government or any of its agencies.

\noindent
\section{Data availability}
The data that support the findings of this article are openly available \cite{github}.

\section{Appendix A: Comparison with geostationary satellites}

\begin{figure}[b]
    \centering
    \includegraphics[width=0.99\linewidth]{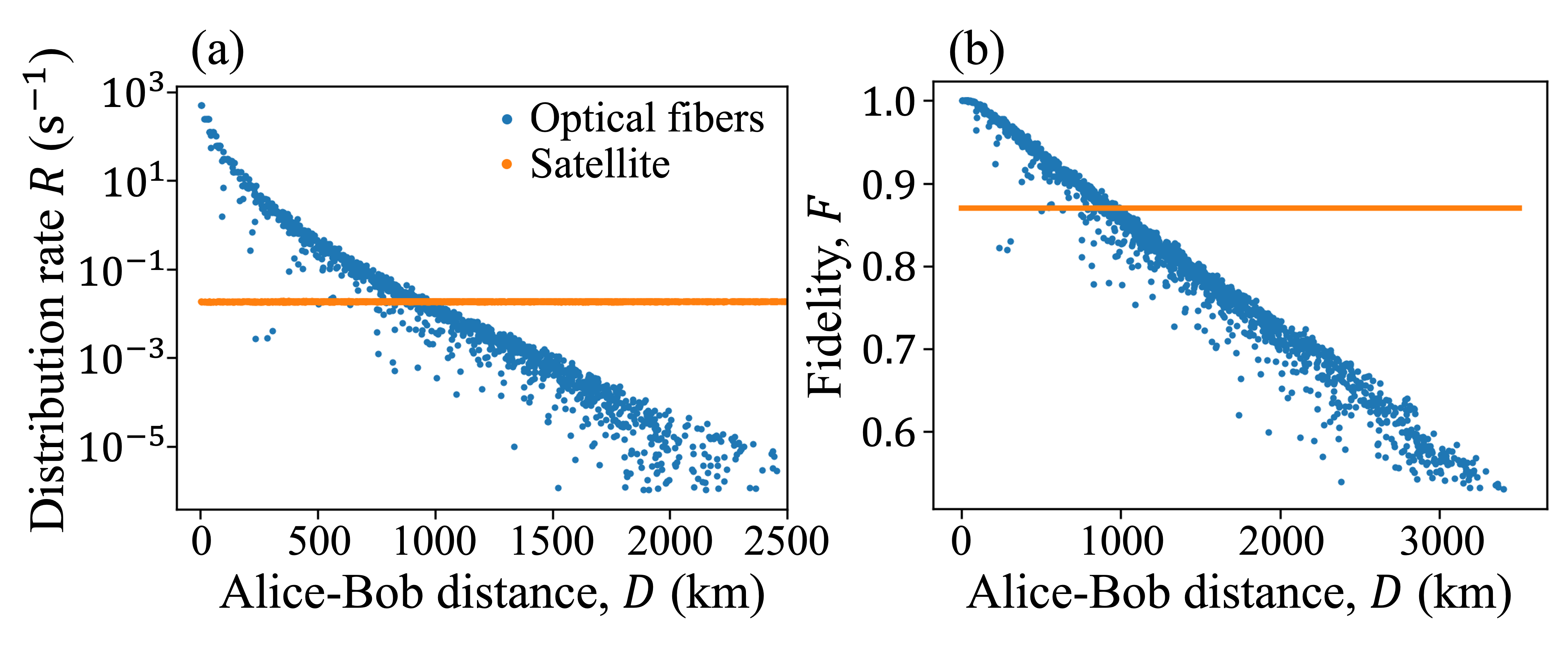}
    \caption{Performance comparison for entanglement distribution protocols using optical fibers versus GEO satellites.  (a) Entanglement distribution rate for optical fibers (blue) and a GEO satellite (orange) with a target fidelity of 0.87. 
    (b) End-to-end entanglement fidelity within the time needed for the satellite to establish one entanglement---the color code is the same as in (a). }
    \label{fig:fiber_satellite_comparison}
\end{figure}

An alternative to MEO satellites would be GEO satellites,
and it is therefore instructive to compare the results for the two different systems.
GEO satellites orbit Earth approximately $36\,000$ km above the equator and remain nearly stationary relative to a fixed point on the ground. This geosynchronous positioning, explored in classical telecommunications, simplifies the operation of ground 
stations, including satellite tracking and pointing. Additionally, the high altitude of the GEO satellites results in a large coverage of the Earth's surface, enabling communication between distant locations with a single satellite.

\begin{figure}
    \centering
    \includegraphics[width=0.99\linewidth]{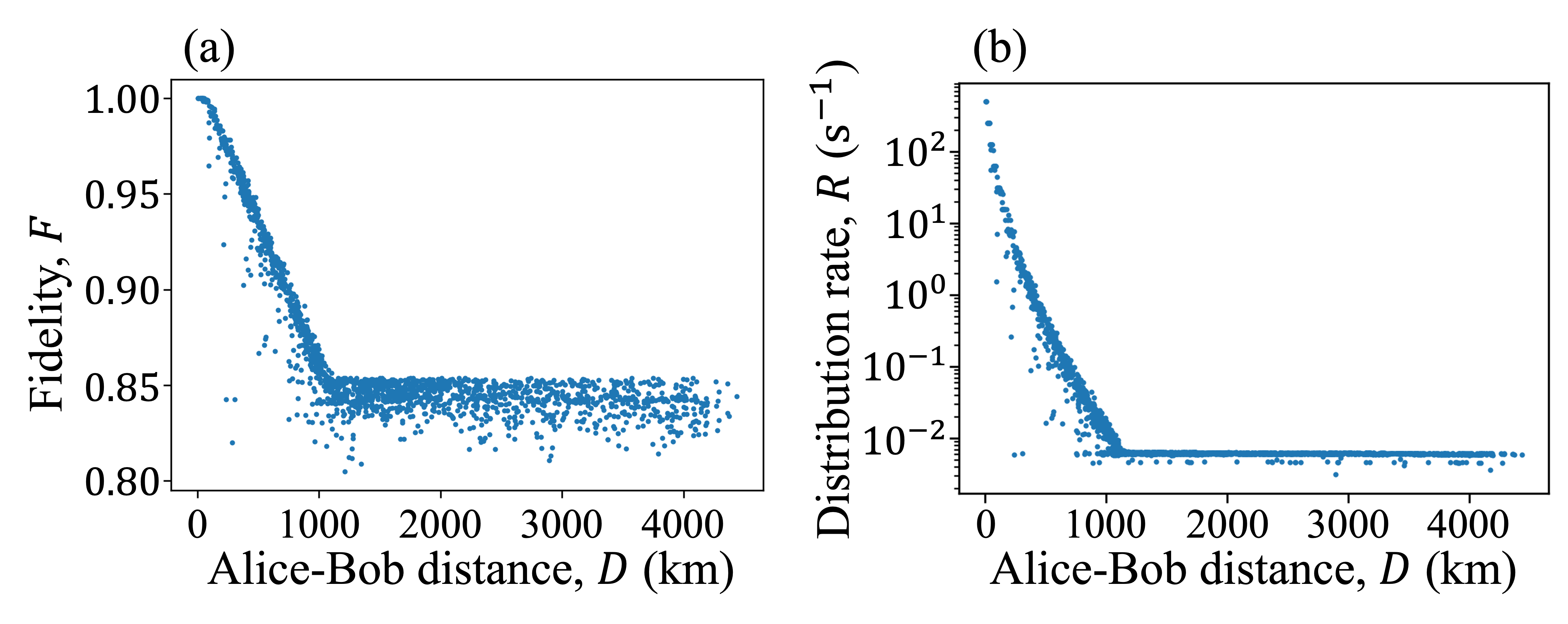}
    \caption{Performance of the hybrid protocol integrating optical fibers and a GEO satellite. (a) End-to-end entanglement fidelity within the time needed to establish one entanglement through the satellite. (b) Entanglement distribution rate of the hybrid protocol with a target fidelity of 0.87.  
    }
    \label{fig:geo_hybrid}
\end{figure}

We assume that each GEO satellite has multiple entangled photon generators and emits $N_{\text{s}}=6\times 10^7$ total pairs of entangled photons per second, as assumed for the MEO satellites. Using a single satellite for the contiguous portion of the United States, the satellite can be positioned at the longitude that minimizes the average satellite-ground distance. This results in an approximate entanglement distribution rate of only $0.02$ s$^{-1}$, which is almost independent of the distance between the ground stations
[Fig.~\ref{fig:fiber_satellite_comparison}(a)]. For reference, the rate decreases rapidly over distance and becomes lower for distances above $950$ km 
if only optical fibers are used with the specifications assumed throughout the article. 
Here, too, the fidelity of entanglement distributed by the satellite is constant, while the fidelity of entanglement distributed by optical fibers decreases with distance [Fig.~\ref{fig:fiber_satellite_comparison}(b)]. 
As shown in Fig.~\ref{fig:geo_hybrid}, the performance of the hybrid protocol combining optical fibers and a GEO satellite is significantly lower than for the corresponding hybrid protocol using MEO satellites. This is due to the higher loss of photons resulting from the higher altitude of these satellites.

\section{Appendix B: Varying the Entanglement-swapping fidelity}
\begin{figure}
    \centering\includegraphics[width=0.85\linewidth]{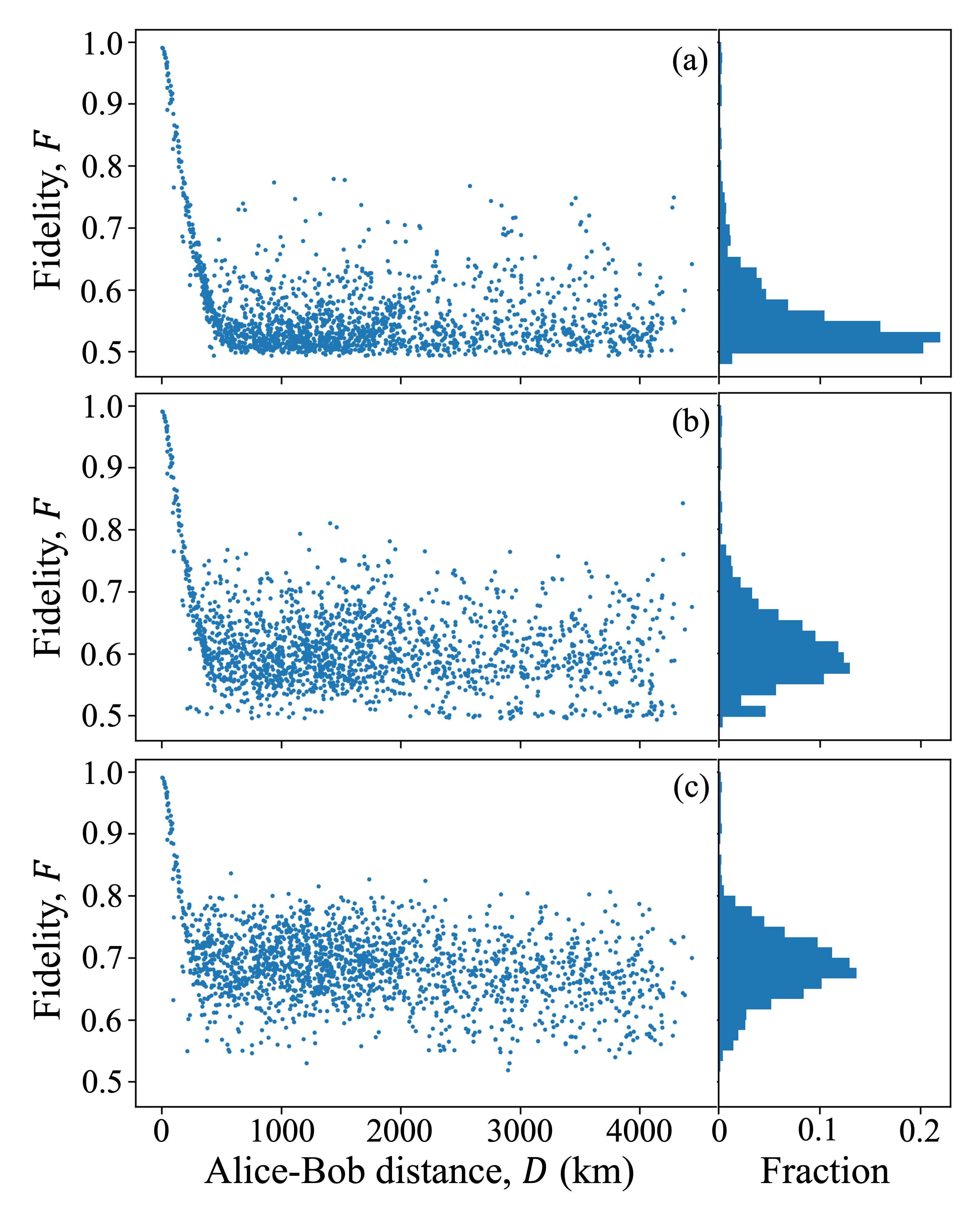}
    \caption{End-to-end entanglement fidelity of the hybrid protocol in the main text for $F_{\text{swap}}^{\text{p}}=0.9$.
    (a-c) Scattered plot (left) and histogram (right) of the end-to-end fidelity using
    increasingly fine grid sizes for the satellite ground stations (top to bottom). The grid sizes  are the same as in Fig.~6(a)-(c). }
    \label{fig:F_swap=0.9}
\end{figure}

Throughout the article, we assume nearly ideal hardware, where the fidelities of generated trapped-ion photon entanglement and entanglement swapping are $F_0$=$F_{\text{swap}}^{\text{p}}$=$F_{\text{swap}}^{\text{i}}=0.99$. 
However, the current technology of entanglement creation and Bell-state measurement renders significantly lower fidelities, especially for photon entanglement swapping. Here, we test the impact of lowering $F_{\text{swap}}^{\text{p}}$.
Figure~\ref{fig:F_swap=0.9} shows the fidelity of end-to-end entanglement established by the hybrid protocol when $F_{\text{swap}}^{\text{p}}=0.9$ and the other parameters remain unchanged. Compared with $F_{\text{swap}}^{\text{p}}=0.99$ in Fig.~6, the end-to-end entanglement fidelity is necessarily  lower 
for $F_{\text{swap}}^{\text{p}}=0.9$
but the average remains above $0.5$. For the satellite ground stations positioned using the same grid sizes, the median end-to-end fidelity changes from $0.71$, $0.76$, and $0.80$ in Fig.~6 to  $0.54$, $0.60$, and $0.69$ in Fig.~\ref{fig:F_swap=0.9}.\\

\section{Appendix C: Varying the ground station telescope size}
\begin{figure}
\centering
\includegraphics[width=0.85\linewidth]{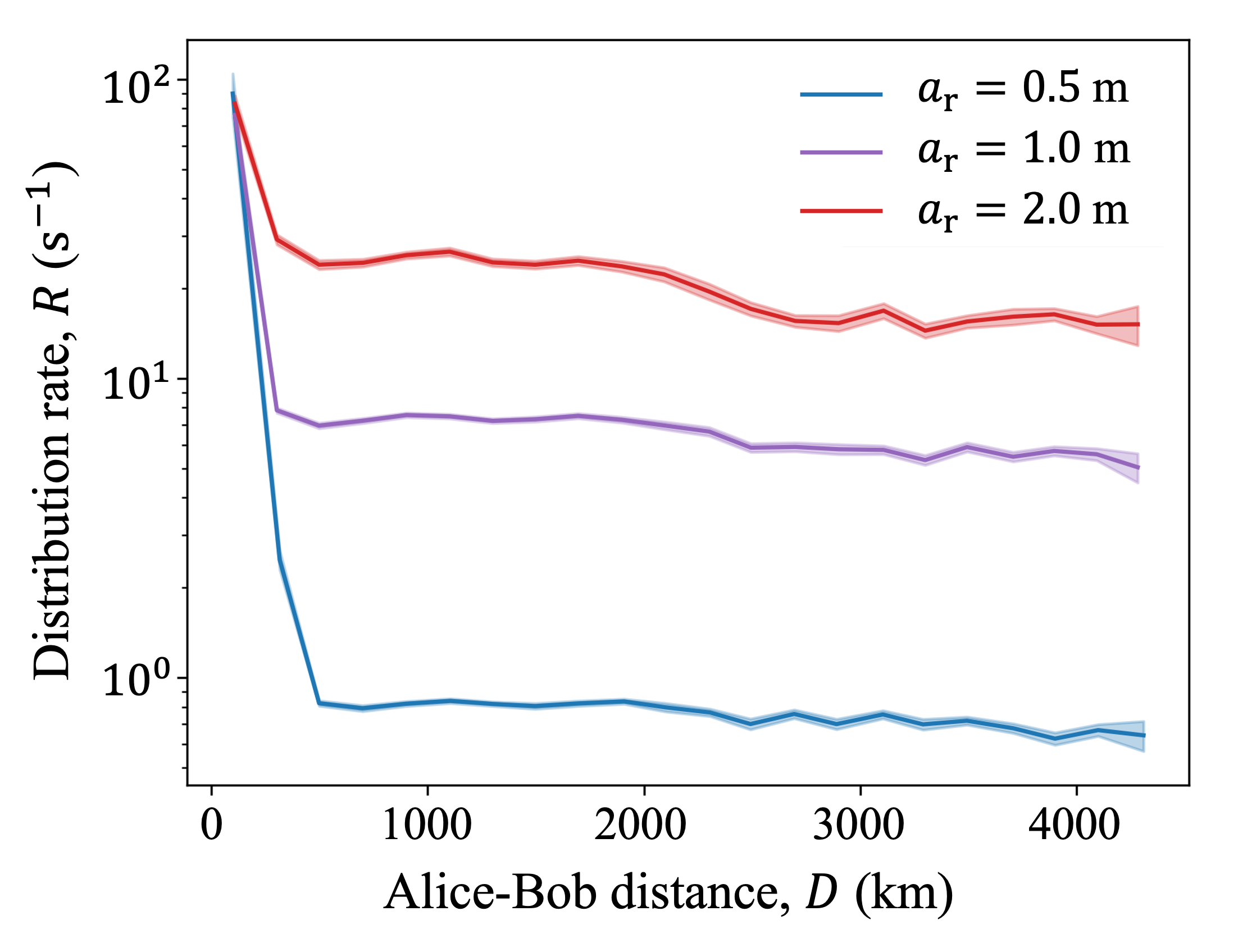}
    \caption{End-to-end entanglement distribution rate for ground-station telescopes of increasing diameter: $a_{\text r}=0.5$ m (blue), $1.0$ m (purple), and $2.0$ m (red). The data are binned to facilitate comparison across cases, with each point representing an average and the shaded bands representing the standard errors. The case $a_{\text r}=0.5$ m corresponds to the scattered plot in Fig.~6(i).
    }
    \label{fig:aR}
\end{figure}

The entanglement distribution rate through the satellite is proportional to the square of the radius of the ground-station telescope, as expressed in Eq.~(6). Therefore, increasing the telescope radius can further improve the performance of the hybrid entanglement distribution protocol, as shown in Fig.~\ref{fig:aR}. 
We emphasize that the target end-to-end fidelity does not depend on the entanglement distribution rate (and thus on the telescope size).

\section{Appendix D: Impact of the decoherence time}
\begin{figure}[t]
    \centering\includegraphics[width=0.85\linewidth]{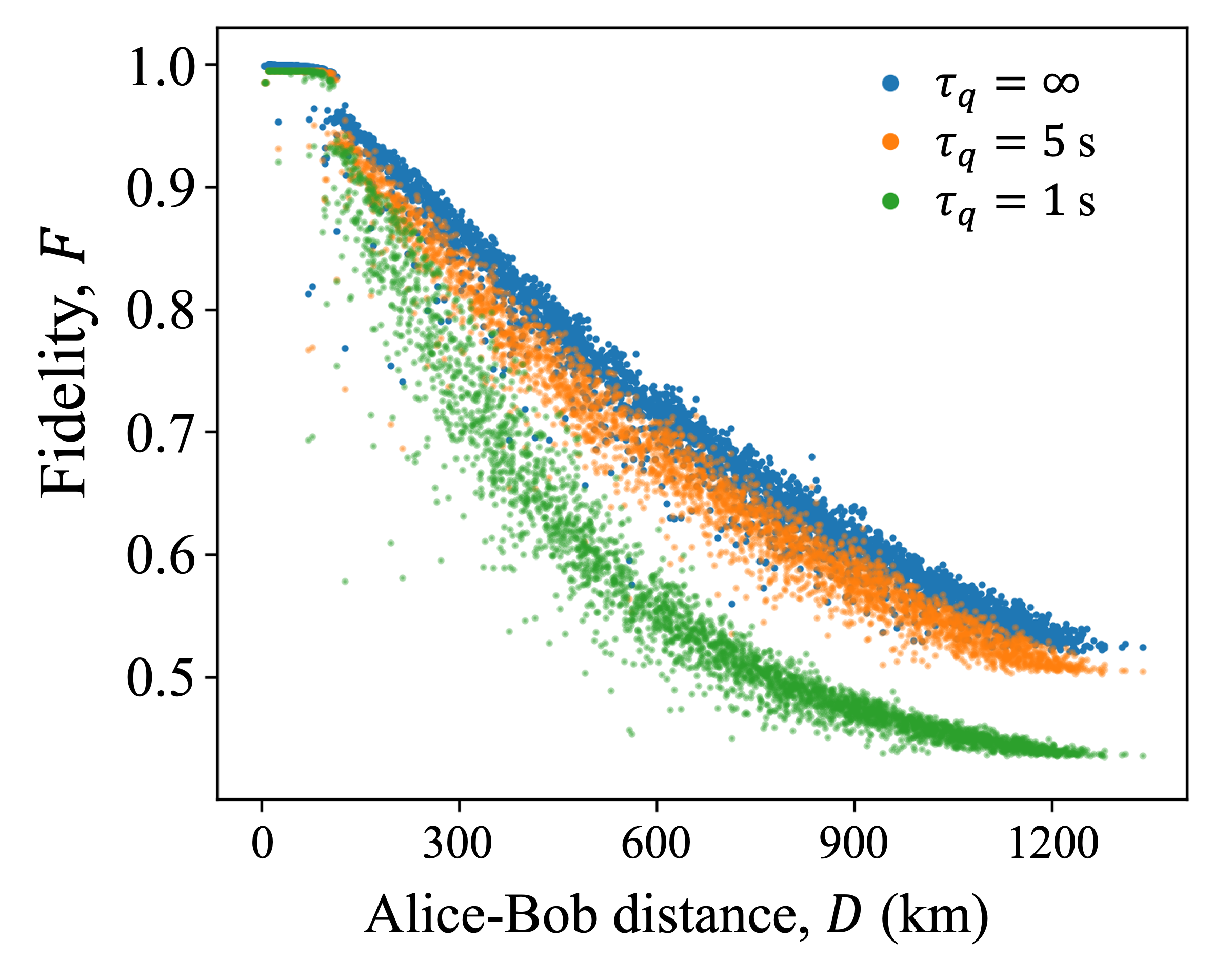}
    \caption{End-to-end entanglement fidelity via optical fiber when fidelity reduction is considered for different decoherence times: $\tau_q=\infty$ [blue, same as in Fig. 4(b)], $\tau_q=5$ s (orange), and $\tau_q=1$~s (green). }
    \label{fig:decoherence}
\end{figure}

Our main results are presented assuming that the decoherence in the qubits stored in quantum memories is negligible. Here, we explicitly investigate the impact of decoherence on the end-to-end fidelity degradation. As described in Eq.~(4), we model decoherence using dephasing noise. In this model, an arbitrary initial state $\rho_0$ evolves as
\begin{equation*}
    {\rho}(t) = q(t) \rho_0 + [1-q(t)]S_z^\dagger \rho_0 S_z, 
\end{equation*}
where $q(t)=(1-e^{-t^2/\tau_q^2})/2$ and $\tau_q$ is the decoherence time. 
After entanglement is established between neighboring trapped-ion repeaters, the qubits are stored in quantum memories while waiting for the next link to complete entanglement purification. 
Figure~\ref{fig:decoherence} shows, for different values of $\tau_q$,  the resulting end-to-end fidelity between randomly selected Alice-and-Bob pairs using only optical fibers as in Fig.~4(b). Shorter decoherence times lead to decreased performance of the fiber-based protocol, as expected. However, the end-to-end fidelity for $\tau_q$ as small as $5$ s is comparable to that for $\tau_q=\infty$.

\section{Appendix E: Impact of the Satellite location}
\begin{figure}[t]
\centering\includegraphics[width=0.99\linewidth]{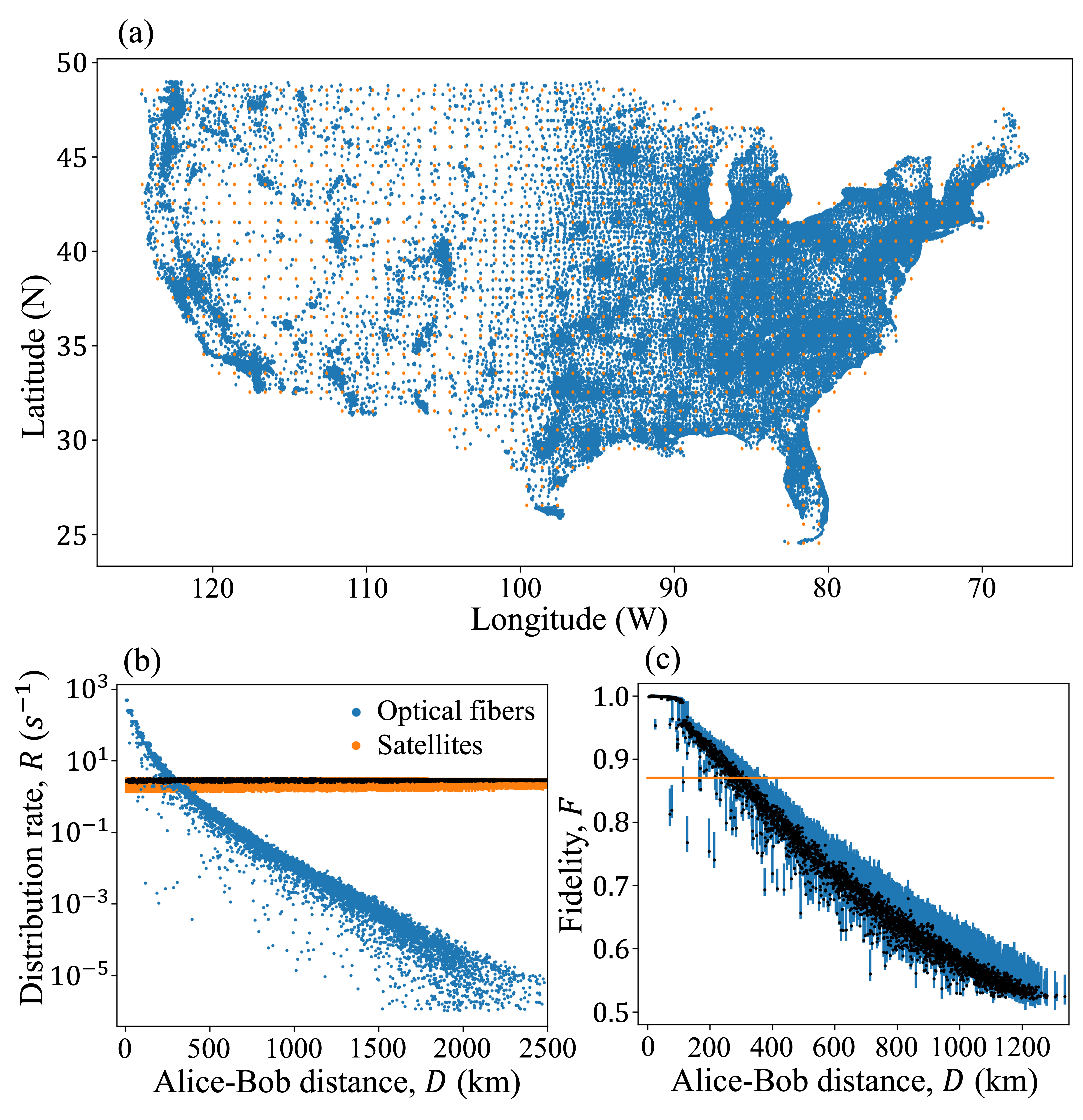}
    \caption{End-to-end entanglement distribution rate and fidelity for different MEO satellite locations. (a) Grid of different MEO subsatellite locations (orange dots)  
    across the contiguous U.S. and near its borders. (b)~Distribution rate through optical fibers (blue) and satellites (orange), where the latter considers the different satellite locations in (a) for each Alice-Bob pair. The black dots indicate the corresponding results from Fig.~4(a),  where the satellite location is  approximated to be at the center of the U.S. (c)  Corresponding fidelity of the end-to-end entanglement for the time needed to establish one entanglement, where the color code is the same as in (b). The small differences between these and the results in Fig.~4 are due to the dependence of diffraction and atmospheric extinction on the satellite's position. 
    }
    \label{fig:sat_loc}
\end{figure}

We assume that at all times there is at least one MEO satellite above the contiguous United States available for entanglement distribution. Because the satellite location changes iover time, here we evaluate the impact of satellite location on entanglement distribution efficiency. Specifically, we sample a grid of subsatellite locations for a grid spacing of 1$^\text{o}$ in both latitude and longitude. The subsatellite locations are placed within or near the U.S. border [orange dots in Fig.~\ref{fig:sat_loc}(a)] so that every node in the network has at least one subsatellite point within a circle of diameter $\sqrt{2}\times 1^\text{o}$ centered on it. 
For each sampled location, we calculate the corresponding entanglement distribution rates and end-to-end fidelity for the same $10\,000$ Alice-Bob pairs considered in Fig.~4. Figure \ref{fig:sat_loc}(b)-(c) shows the resulting distribution rates and fidelities across  satellite locations via optical fibers (blue) or satellites (orange), compared with the case in which we approximate the satellite location as being at the center of the map (black). 
These results indicate that, provided the satellite is positioned above the contiguous U.S., its exact location has only a small effect on the performance of the entanglement distribution protocol,  supporting the validity of the satellite-location approximation used throughout the article. 

\begin{thebibliography}{99}


\bibitem{gisin2007quantum}
N. Gisin and R. Thew, Quantum communication, Nat. Photonics \textbf{1}, 165 (2007). 

\bibitem{chung2021illinois}
J. Chung \textit{et al.}, Illinois Express Quantum Network (IEQNET): metropolitan-scale experimental quantum networking over deployed optical fiber, Proc. SPIE Int. Soc. Opt. Eng. \textbf{11726}, 1172602 (2021). 

\bibitem{bersin2024development}
E. Bersin \textit{et al.}, Development of a Boston-area 50-km fiber quantum network testbed, Phys. Rev. Appl. \textbf{21}, 014024 (2024). 

\bibitem{martin2023madqci}
V. Martin \textit{et al.}, MadQCI: a heterogeneous and scalable SDN-QKD network deployed in production facilities, npj Quantum Inf. {\bf 10}, 80 (2024). 

\bibitem{garcia2024strategies}
M.I. Garcia-Cid, L. Ortiz, J. Saez, and V. Martin, Strategies for the Integration of quantum networks for a future quantum internet, arXiv:2401.06444 (2024).

\bibitem{chen2021integrated}
YA. Chen \textit{et al.}, An integrated space-to-ground quantum communication network over 4,600 kilometres, Nature \textbf{589}, 214–219 (2021). 

\bibitem{xu2020secure}
F. Xu, X. Ma, Q. Zhang, H.K. Lo, and J.W. Pan, Secure quantum key distribution with realistic devices, Rev. Mod. Phys. \textbf{92}, 025002 (2020).

\bibitem{cao2022evolution}
Y. Cao, Y. Zhao, Q. Wang, J. Zhang, S. X. Ng, and L. Hanzo, The Evolution of Quantum Key Distribution Networks: On the Road to the Qinternet, IEEE Communications Surveys \& Tutorials \textbf{24}, 839 (2022). 

\bibitem{cacciapuoti2020quantum}
A. S. Cacciapuoti, M. Caleffi, F. Tafuri, F. S. Cataliotti, S. Gherardini, and G. Bianchi, Quantum Internet: Networking Challenges in Distributed Quantum Computing, IEEE Network \textbf{34}(1), 137 (2020).

\bibitem{sanchez-burillo2012quantum}
E. Sánchez-Burillo, J. Duch, J. Gómez-Gardenes, and D. Zueco, Quantum Navigation and Ranking in Complex Networks, Sci. Rep. \textbf{2}, 605 (2012).

\bibitem{pirandola2018advances}
S. Pirandola, B. R. Bardhan, T. Gehring, C. Weedbrook, and S. Lloyd, Advances in photonic quantum sensing, Nature Photon. \textbf{12}, 724 (2018).

\bibitem{kimble2008quantum}
H. Kimble, The quantum internet, Nature \textbf{453}, 1023 (2008).

\bibitem{wehner2018quantum}
S. Wehner, D. Elkouss, and R. Hanson, Quantum internet: A vision for the road ahead, Science \textbf{362}, 9288 (2018).

\bibitem{wengerowsky2019entanglement}
S. Wengerowsky, S.K. Joshi, F. Steinlechner, J.R. Zichi, S.M. Dobrovolskiy, R. Van der Molen, J.W. Los, V. Zwiller, M.A. Versteegh, A. Mura, and D. Calonico, Entanglement distribution over a 96-km-long submarine optical fiber, Proc. Natl. Acad. Sci. USA \textbf{116}(14), 6684 (2019).

\bibitem{neumann2022continuous}
S.P. Neumann, A. Buchner, L. Bulla, M. Bohmann, and R. Ursin, Continuous entanglement distribution over a transnational 248 km fiber link, Nat. Commun. \textbf{13}, 6134 (2022). 

\bibitem{briegel1998quantum}
H.-J. Briegel, W. Dür, J. I. Cirac, and P. Zoller, Quantum repeaters: the role of imperfect local operations in quantum communication, Phys. Rev. Lett. \textbf{81}, 5932 (1998). 

\bibitem{azuma2023quantum}
K. Azuma, S.E. Economou, D. Elkouss, P. Hilaire, L. Jiang, H. Lo, and I. Tzitrin, Quantum repeaters: From quantum networks to the quantum internet, Rev. Mod. Phys. \textbf{95}, 045006 (2023).

\bibitem{avis2023requirements}
G. Avis, F. Ferreira da Silva, T. Coopmans, A. Dahlberg, H. Jirovská, D. Maier, J. Rabbie, A. Torres-Knoop, and S. Wehner, Requirements for a processing-node quantum repeater on a real-world fiber grid, npj Quantum Inf. \textbf{9}, 100 (2023). 

\bibitem{vallone2015experimental}
G. Vallone, D. Bacco, D. Dequal, S. Gaiarin, V. Luceri, G. Bianco, and P. Villoresi, Experimental Satellite Quantum Communications, Phys. Rev. Lett. \textbf{115}, 040502 (2015). 

\bibitem{yin2017satellite}
J. Yin, \textit{et al.}, Satellite-based entanglement distribution over 1200 kilometers, Science \textbf{356}, 1140 (2017).

\bibitem{lu2022micius}
C.Y. Lu, Y. Cao, C.Z. Peng, and J.W. Pan, Micius quantum experiments in space, Rev. Mod. Phys. \textbf{94}, 035001 (2022). 

\bibitem{liao2018satellite}
S. Liao \textit{et al.}, Satellite-Relayed Intercontinental Quantum Network, Phys. Rev. Lett. \textbf{120}, 030501 (2018).

\bibitem{williams2024scalable}
A. Williams, N.K. Panigrahy, A. McGregor, and D. Towsley, Scalable Scheduling Policies for Quantum Satellite Networks, in 2024 IEEE International Conference on Quantum Computing and Engineering (QCE), (IEEE, 2024), pp. 1760–1769.

\bibitem{wengerowsky2018entanglement}
S. Wengerowsky, S.K. Joshi, F. Steinlechner {\it et al}., An entanglement-based wavelength-multiplexed quantum communication network, Nature {\bf 564}, 225 (2018).

\bibitem{miao2007feasibility}
E.L. Miao, Z.F. Han, T. Zhang, and G.C. Guo, The feasibility of geostationary satellite-to-ground quantum key distribution, Phys. Lett. A \textbf{361}, 29 (2007).

\bibitem{vrutyanskiy2023telecom}
V. Krutyanskiy, M. Canteri, M. Meraner, J. Bate, V. Krcmarsky, J. Schupp, N. Sangouard, and B.P. Lanyon, Telecom-wavelength quantum repeater node based on a trapped-ion processor, Phys. Rev. Lett. \textbf{130}, 213601 (2023). 

\bibitem{sangouard2009quantum}
N. Sangouard, R. Dubessy, and C. Simon, Quantum repeaters based on single trapped ions, Phys. Rev. A \textbf{79}, 042340 (2009). 

\bibitem{meraner2020indistinguishable}
M. Meraner, A. Mazloom, V. Krutyanskiy, V. Krcmarsky, J. Schupp, D. A. Fioretto, P. Sekatski, T. E. Northup, 
N. Sangouard, and B. P. Lanyon, Indistinguishable photons from a trapped-ion quantum network node, Phys. Rev. A \textbf{102}, 052614 (2020).

\bibitem{drmota2023robust}
P. Drmota \textit{et al.} Robust Quantum Memory in a Trapped-Ion Quantum Network Node, Phys. Rev. Lett. \textbf{130}, 090803 (2023).

\bibitem{noteSM1}
This formulation anticipates improvements over current technologies for $F_0$, $F_{\text{swap}}^{\text{p}}$, and $F_{\text{swap}}^{\text{i}}$  \cite{meraner2020indistinguishable}. Among these, $F_{\text{swap}}^{\text{p}}$ remains the most limiting factor, and its impact is examined in Appendix B. 

\bibitem{wang2019efficient}
Y. Wang, J. Li, S. Zhang, K. Su, Y. Zhou, K. Liao, S. Du, H. Yan, and S. Zhu, Efficient quantum memory for single-photon polarization qubits, Nat. Photonics \textbf{13}, 346 (2019).  

\bibitem{pirandola2021satellite}
S. Pirandola, Satellite quantum communications: fundamental bounds and practical security, Phys. Rev. Res. \textbf{3}, 023130 (2021). 


\bibitem{gruneisen2017modeling}
M.T. Gruneisen, M.B. Flanagan, and B.A. Sickmiller, Modeling satellite-Earth quantum channel downlinks with adaptive-optics coupling to single-mode fibers, Opt. Eng. {\bf56}(12), 126111 (2017). 

\bibitem{gruneisen2021adaptive}
M.T. Gruneisen \textit{et al.}, Adaptive-Optics-Enabled Quantum Communication: A Technique for Daytime Space-To-Earth Links, Phys. Rev. Applied {\bf 16}, 014067 (2021). 

\bibitem{cameron2024adaptive}
P. Cameron, B. Courme, C. Verni\`ere, R. Pandya, and H. Defienne, Adaptive optical imaging with entangled photons, Science {\bf383}, 1142 (2024).

\bibitem{scarfe2025fast}
L. Scarfe, F. Hufnagel, M.F. Ferrer-Garcia, A. D’Errico, K. Heshami, and E. Karimi, Fast adaptive optics for high-dimensional quantum communications in turbulent channels, Commun. Phys. {\bf8}, 79 (2025).


\bibitem{census2020}
2020 Census National Redistricting Data Summary File, Prepared by the U.S. Census Bureau, 2021. 


\bibitem{goswami2023satellite}
S. Goswami and S. Dhara, Satellite-relayed global quantum communication without quantum memory, Phys. Rev. Applied \textbf{20}, 024048 (2023). 

\bibitem{bacciottini2024leveraging}
L. Bacciottini, A. Chandra, M.G. De Andrade, N.K. Panigrahy, S. Pouryousef, N.S. Rao, E. Van Milligen, G. Vardoyan, and D. Towsley, Leveraging Internet Principles to Build a Quantum Network, arXiv:2410.08980 (2024).

\bibitem{github}
The data and associated code are available at the following GitHub repository: \url{https://github.com/ShaoYanxuan/hybrid_quantum_network.git}. 

\end{thebibliography}
\end{document}